\documentclass[amsmath,amssymb,aps,prd,nofootinbib,eqsecnum,a4paper,superscriptaddress]{revtex4} 

\usepackage{amsmath}
\usepackage{amsfonts}
\usepackage{amssymb}
\usepackage{graphicx}
\usepackage{psfrag}

\usepackage{graphicx,bm}
\def\be{\begin{equation}}
\def\ee{\end{equation}}
\def\bea{\begin{eqnarray}}
\def\eea{\end{eqnarray}}
\def\nn{\nonumber}
\def \ni {\noindent}
\def\drho{{\delta\rho_1}}
\def\drhorho{{\delta\rho_2}}

\def\dP{{\delta P_1}}

\def\ipsi{{\psi_1}}
\def\iipsi{{\psi_2}}
\def\E{{E_1}}
\def\EE{{E_2}}
\def\iphi{{\phi_1}}
\def\iiphi{{\phi_2}}
\def\B{{B_1}}
\def\BB{{B_2}}

\def\dij{\delta_{ij}}

\def \beg {\begin{enumerate}}
\def \en {\end{enumerate}}

\def\Pb{P_0}
\def\rhob{\rho_0}

\def\cs{c_{\rm{s}}^2}

\def\wt{\widetilde}
\def\n{\nabla}

\def\H{{\cal H}}
\def\cs2{c_{\rm{s}}^2}

\def\U0{{\bar U_0}}
\def\wt{\widetilde}
\def\dij{\delta_{ij}}
\def\bi{\begin{itemize}}
\def\ei{\end{itemize}}

\newcommand\New{\mathrm{N}}

\def\U{{\cal{U}}}

\def\fNL{f_{\rm NL}}

\def\tom{{\rm{tom}}}

\def\com{{\rm{com}}}

\def\lg{{\ell}}

\usepackage[usenames]{color}
\usepackage{psfrag}
\definecolor{myred}{rgb}{0.85,0.08,0}
\definecolor{mydb}{rgb}{0,0.08,0.8}

\usepackage{hyperref}

\begin{document}

\title{The Poisson equation at second order in relativistic cosmology}  

\author{J.~C.~Hidalgo}
\email[]{juan.hidalgo@port.ac.uk}
\affiliation{Institute of Cosmology and Gravitation, University of Portsmouth,
 Portsmouth, PO1 3FX, UK}
\affiliation{Instituto de Astronom\'ia, UNAM, Ciudad Universitaria,
  04510, M\'exico D.F., M\'exico}

\author{Adam J.~Christopherson}
\email[]{Adam.Christopherson@nottingham.ac.uk}
\affiliation{School of Physics and Astronomy, University of
  Nottingham, University Park, Nottingham, NG7 2RD, UK}

\author{Karim~A.~Malik} 
\email[]{k.malik@qmul.ac.uk}
\affiliation{Astronomy Unit, School of Physics and Astronomy, Queen
  Mary University of London,Mile End Road, London, E1 4NS, UK}

\date{\today}

\begin{abstract}
  We calculate the relativistic constraint equation which relates the
  curvature perturbation to the matter density contrast at second
  order in cosmological perturbation theory. This relativistic
  ``second order Poisson equation'' is presented in a gauge where the
  hydrodynamical inhomogeneities coincide with their Newtonian
  counterparts exactly for a perfect fluid with constant equation of
  state. We use this constraint to introduce primordial
  non-Gaussianity in the density contrast in the framework of General
  Relativity. We then derive expressions that can be used as the
  initial conditions of N-body codes for structure formation which
  probe the observable signature of primordial non-Gaussianity in the
  statistics of the evolved matter density field.
\end{abstract}


\maketitle

\section{Introduction}

Our knowledge of the statistics of the galaxy distribution relies upon the vast
amount of data obtained by increasingly large galaxy surveys
\cite{Hawkins:2002sg, Eisenstein:2011sa, Schelgel:2011zz, Laureijs:2011mu}.
Among other goals, analysis of the galaxy field allows us to
indirectly probe the distribution of the underlying dark matter on non-linear
scales (e.g.,Refs.~\cite{peebles:80,Bernardeau:2001qr}). 
On the theoretical side, in order to understand the physics that
governs the observed galaxy field, large numerical codes are developed
to simulate the evolution of matter inhomogeneities that have formed
large scale structure (LSS). This huge task is usually split in two
stages. In a first stage, semi-analytical methods are employed to
account for the early evolution of fluctuations in the weakly
non-linear regime. At the same time, the inhomogeneous in the
continuum matter field are related to a discrete distribution of point
masses, thus implementing initial conditions for numerical codes. In a
second stage, typically at redshifts $z\sim 50$,  N-body codes evolve
inhomogeneities in the strongly non-linear regime up to the present
day. As Newtonian N-body codes continue to improve in resolution and
volume (e.g., Refs.~\cite{Evrard:2001hu,Crocce:2009mg,Habib:2012qs}),
the implementation of realistic and accurate initial conditions is
increasingly important.   
 
Historically, the initial conditions for N-body simulations have been
generated by using the Zel'dovich approximation
\cite{Zeldovich:1969sb}, which establishes the correspondence between
the matter density fluctuation of standard perturbation theory, and
the displacement of mass particles in a grid. Despite its linear
nature, this represents an improvement over standard perturbation
theory, since it takes advantage of working in Lagrangian coordinates
\cite{Bernardeau:2001qr,Yoshisato:2005ks}. 
The caveat to this approximation is that it accounts only for the early
non-linear evolution of density fluctuations, and in particular, it employs a
linear Poisson constraint, which is used to express the density contrast,
$\delta_{\rm N}$, in terms of the gravitational potential, 
$\phi_{\rm N}$, that is
\be
\nabla^2 \phi_{\New} = 4 \pi G \rhob a^2 \delta_{\New}.
\label{newton:poisson}
\ee

An improvement to this approximation is second-order Lagrangian
perturbation theory (2LPT), which generates initial 
conditions taking into account non-linearities in Lagrangian
coordinates. This has been shown to be more precise and avoids
transients present in the Zel'dovich approximation
\cite{Crocce:2006ve,Sirko:2005uz}.   
Since 2LPT takes into account non-linearities, the fact that the
gravitational instability is non-local is manifest in corrections to
Eq.~(\ref{newton:poisson}) given by tidal effects at non-linear order
\cite{Buchert:1993df}.
With the matter density fluctuations at non-linear order under
control, recent studies have used 2LPT to include primordial non-Gaussianity
in the matter fluctuations
\cite{Dalal:2007cu,Desjacques:2008vf,Scoccimarro:2011pz}.     

These and other semi-analytical approximations to the early
evolution of inhomogeneities, however, rely on Newtonian physics,
thereby ignoring the effects of General Relativity (GR). Cosmological
inhomogeneities are well described by Newtonian dynamics only when the
modes of the perturbations lie well inside the horizon, i.e.~when
their wavenumber is $k\gg \H$, with $\H$ denoting the Hubble parameter
in conformal time. Yet, the initial conditions for these
approximations come from much earlier times -- typically the epoch of
decoupling -- when some of the scales of interest are comparable to,
or even larger than, the cosmological horizon. Therefore, relativistic
effects are important and should be taken into account when setting the
initial conditions to simulations of structure formation. 

Recent studies demonstrate the importance of GR in the analysis of
large scale structure. 
Some have contrasted relativistic and Newtonian fluctuations
by the identification of dynamical equations
\cite{Chisari:2011iq,Green:2011wc}. This provides
correspondences between Newtonian fluctuations and relativistic
perturbations in a specific gauge at linear order in perturbation
theory. Additionally, Ref.~\cite{Hwang:2012bi} extends this
correspondences to second-order perturbations. In this way, the equivalence of the dynamical equations
is established for the restricted case of pressureless matter and
neglecting the decaying mode of perturbations.  
A major motivation to study this correspondence is to
discriminate primordial non-Gaussian fluctuations from
non-Gaussianities induced by the non-linear dynamics of GR. In
search of observational signatures,  
Ref.~\cite{Bruni:2011ta} studied the effects of relativistic non-linear
fluctuations in the halo bias and subsequently its signature in the
spectrum of the galaxy distribution (see also \cite{Desjacques:2010jw}).  
\\

In this paper we present the Poisson equation at second order in the
framework of relativistic cosmological perturbation theory
\cite{Bardeen:1980kt, mfb, ks, MW2008, MM2008}. 
Previous studies have explored this constraint for the limit of
a dust universe at small scales \cite{Hwang:2012bi} (in this case the linear
equation \eqref{newton:poisson} is recovered), and for a
$\Lambda$CDM universe at large scales \cite{Bartolo:2010rw}. Instead,
our analysis yields the Poisson constraint equation in terms of
relativistic perturbations that find a 
direct correspondence with Newtonian inhomogeneities, and without
approximations. Furthermore, we extend the constraint to the case of a
general perfect fluid.  
As an example, we subsequently use our result to express the primordial
non-Gaussianity in terms of the dark matter density field
in equations that include all the relativistic effects.
We present results in the form of kernels for the non-linear variables,
a form customarily used in the formulation of initial conditions of
numerical simulations.   

The paper is organised as follows. In the next section we
explicitly show how to construct the Poisson equation from
Einstein's field equations combining variables in two gauges for
linear perturbations. 
In Section \ref{sec:second:order} we repeat the
procedure for the second order variables and arrive at a GR version of
the Poisson constraint valid for any perfect fluid including entropy
(or non-adiabatic pressure) perturbations.
In Section \ref{sec:initial:conditions} we apply the
constraint to the case of matter perturbations in a flat universe
dominated by pressureless matter and show how to include the
primordial non-Gaussian 
corrections in the Poisson equation. We conclude in Section
\ref{sec:discussion} discussing the relevance of our result
to the initial conditions of numerical simulations. 

\section{The Poisson equation at first order}
\label{sec:first:order}

\subsection{Background and first-order equations}

In cosmological perturbation theory, considering
scalar perturbations of the metric yields the following line element,
\be
ds^2=a^2(\eta)\Big\{-(1+2\phi)d\eta^2+2B_{,i}dx^id\eta
+\Big[(1-2\psi)\dij+E_{,ij}\Big]dx^idx^j\Big\},
\label{pert:metric}
\ee  
where $\phi$ is the lapse function, $\psi$ is the
curvature perturbation, and $B$ and $E$ make up the scalar
shear. All these quantities are function of Cartesian coordinates,
${x}_j$, and conformal time, $\eta$. Perturbations are then expanded
order-by-order in a series as, e.g., $\phi=\phi_1+\frac{1}{2}\phi_2+\cdots$. 
In order to define the expansion uniquely, and
as an excellent approximation to observations, the first
order quantities are chosen to have Gaussian statistics. 

In the background the metric represents the
Friedmann-Lema\^itre-Robertson-Walker spacetime. The homogeneous
equations are the familiar Friedmann and continuity equations:
\begin{align}
3 \H^2 &= 8 \pi G a^2 \rhob,
\label{friedmann:eq}\\
\rhob' &= - 3 \H (P_0 + \rhob), 
\label{cont:bg}
\end{align}

\noindent where the prime denotes a derivative with respect to
conformal time and a subscript zero denotes the background,
homogeneous quantities. 

The fluid equations are derived from the vanishing covariant
derivative of the energy momentum tensor\footnote{
We consider the usual perfect fluid energy momentum
tensor of the form $T^{\mu}{}_{\nu}=(\rho+P)u^\mu u_\nu+P\delta^\mu{}_\nu$,
where $u^\mu$ is the fluid four velocity and $P$ and $\rho$ are the pressure and 
energy density, respectively.
}.
At first-order in
perturbation theory, the energy conservation
dictates the evolution of the density perturbation
$\delta \rho_1$,
\be
\delta \rho_1' + 3 \H \left(\delta \rho_1 + \delta P_1\right) =
(\rhob + P_0) \left[3\psi_1' - \n^2 (E_1' + v_1)\right],
\label{en:cons}
\ee

\noindent where $v$ is the scalar velocity potential obtained from the 
spatial part of the fluid four velocity as $u^i=\frac{1}{a}\partial^i v$,
and the energy density and pressure fluctuations are denoted by 
$\delta\rho$ and $\delta P$, respectively. We define the
Laplacian as $\n^2 = \partial_j\partial^j$. Note that no gauge
has been specified here. In order to obtain the corresponding
equation for the evolution of the velocity, we define $V = B + v$,
and write the momentum conservation equation, which at first order is  
\be
V_{1,i}' + \H (1 + c_{\rm s}^2)V_{1,i} + \left[\frac{\delta P_{1}}{P_0 +
    \rhob} + \phi_1\right]_{,i} = 0,
\label{mom:cons} 
\ee

\noindent where we have neglected anisotropic stresses and defined
the adiabatic sound speed as $c_{\rm s}^2 = P_0' / \rhob'$.

The Einstein field equations yield two constraint equations that
are combined to derive the Poisson equation. The $ (0,0)$ component
of these equations yields the energy constraint equation
\be
3\H (\psi_1' + \H \phi_1) - \n^2\Big(\psi_1 + \H(E_1' - B_1)\Big) =
- 4 \pi G a^2 \delta\rho_1.
\label{en:const}
\ee

\noindent The momentum constraint is derived from the $(0,i)$
component:
\be
\psi_{1,i}' + \H \phi_{1,i} = - 4 \pi G a^2 (\rhob + P_0)V_{1,i}. 
\label{mom:const}
\ee
This is the complete set of equations at first order without the gauge
specified. The remaining Einstein equations at this order are related
to the ones above through the Bianchi identities. 

\subsection{Constraint in the longitudinal gauge}

In order to overcome the ambiguity in the coordinate freedom, we must
specify the gauge in the above equations. We work in the
longitudinal or Newtonian gauge \cite{Bardeen:1980kt} to
recover the exact Newtonian equations. This is a
shear-free gauge, specified by setting $E_\lg = B_{\lg} = 0$. The 
absence of anisotropic stresses also guarantees that, in this gauge,
$\psi_{1 \lg} = \phi_{1\lg} $ and Eq.~\eqref{en:const} becomes
\be
- 3 \H ( \phi_{1\ell}' + \H \phi_{1\ell}) + \nabla^2 \phi_{1\ell} = 4
\pi G a^2 \delta\rho_{1\lg}.  
\label{en:constlg}
\ee

\noindent Then, by integrating the overall gradient of the momentum
constraint \eqref{mom:const}, we have 
\be
\phi_{1\ell}' + \H\phi_{1\ell} = - 4 \pi G a^2 (\rhob + P_0) v_{1\lg}, 
\label{mom:constlg}
\ee

\noindent and combining both equations we find the first-order constraint:
\be
\n^2 \phi_{1\ell} = 4 \pi G a^2 (\delta\rho_{1\lg} + \rhob'v_{1\lg}).
\label{long:poisson}
\ee

\subsection{The Newtonian expression}

The combination in parentheses in the linear Poisson
equation~\eqref{long:poisson}, $\delta\rho_{1\lg} + \rhob'v_{1\lg}$,
is in fact equivalent to the density contrast
in two other gauges, as we will now show. 
The transformation between two coordinate systems is parametrised through
the generating vector  
$
\xi_{1\mu} = (\alpha_1,\beta_{1,i})
$,
so that, for example, the density perturbation at linear
order is transformed as  
\be
\wt{\delta \rho_1} = \delta \rho_1 + \alpha {\rhob'}\,.
\label{gauge:trans}
\ee
We can define the total matter gauge (denoted with a subscript tom) 
by a vanishing total momentum at all
orders, i.e.  
\be
{V_{\tom}} = {v_{\tom}} + {B_{\tom}} = 0.
\label{tom:gauge}
\ee

\noindent The transformation rule for $V$ tells us that
\be
\wt{V_{1}} = v_1 + B_1 - \alpha_1\,,
\ee

\noindent so that in the case of a transformation from the
longitudinal to the total matter gauge we have 
\be
\alpha_{1 \tom | \ell} = v_{1 \lg}\,,
\label{alpha1:tom}
\ee
where the notation $X_{\tom | \ell}$ denotes the value of the gauge generation vector component $X$ for the total
matter gauge, evaluated in the longitudinal gauge.
In order to fully specify the total matter gauge (i.e. in order to specify $\beta_{1 \rm tom}$) 
the condition $E_{1 \rm tom}=0$ is taken.
In consequence, the density fluctuation in the total matter gauge is obtained 
in terms of matter perturbations in the longitudinal gauge as
\be
\wt{\delta\rho_{1\tom}} = \delta\rho_{1\lg} + \rhob' v_{1\lg}\,.
\label{delta1:tom}
\ee 

\noindent It is now straightforward to recover the Newtonian form of the
Poisson equation writing
\be
\n^2 \phi_{1\ell} = 4 \pi G a^2 \rhob\; \delta_{1\tom}\,,
\label{poisson1:tom}
\ee

\noindent where the density contrast is, at first order, 
$\delta_{1\tom}  = \delta \rho_{1\tom} / \rhob$, and at second order
$\delta_{2\tom} = \delta \rho_{2 \tom} / \rhob$. 

Alternatively, we can perform a similar transformation and define
the comoving gauge (denoted with a subscript com)
where the three velocity of the fluid vanishes 
$v_{\com} = 0$. Then, imposing orthogonality of the constant time
hypersurfaces to the four velocity, requires $v_{\com} +
B_{\com} = 0$. In this case, one finds that $\alpha_{1\com} =
v_{1\lg}$, just as in the total matter gauge. Thus, the matter
density in the comoving gauge at linear order, $\delta \rho_{1\com}$ 
reproduces the expression in
\eqref{alpha1:tom}.   
The corresponding Poisson equation
\be
\n^2 \phi_{1\ell} =4 \pi G a^2 \rhob\,\delta_{1 \com}\,,
\label{poisson1:com}
\ee

\noindent has been recovered in previous works
\cite{peebles:80,Wands:2009ex}. It has further been 
shown that with the same combination of variables (namely
$\delta_{1\com},\, \phi_{1\ell}, \, v_{1\lg}$) one can reproduce the
equations used in Newtonian hydrodynamics at linear order
\cite{Chisari:2011iq,Green:2011wc}, with the exception of fluids with
non-vanishing pressure, and which allow for
entropy perturbations \cite{Christopherson:2012kw}.

However, while at first order the gauge transformation from
the longitudinal gauge into both the total matter and comoving gauges
requires only knowledge of the temporal component of the gauge generating 
vector, at second order we require the spatial component scalar $\beta_1$.
In particular, the gauge transformation $\delta_{2\ell} \to
\delta_{2\com}$ includes time integrals in $\beta_{1 \com}$ which may
introduce non-local terms. Therefore, in this work we avoid this
additional complication by working with the density fluctuation in the
total matter gauge.   

Indeed, constructing the scalar $\beta_{1\tom}$, which is
determined by the transformation $E_{1\tom} = E_{1\lg} + \beta_{1
  \tom} = 0$, we note that it does not involve a time integral. Explicitly 
\be 
\beta_{1\tom|\ell} = 0\,.
\label{beta1:tom}
\ee

\section{The constraint at second order}
\label{sec:second:order}

In the previous section we have shown how the linear energy and momentum
constraint equations can be combined to obtain a Poisson equation at
first order. The same procedure can be followed to write a
Poisson-like constraint at second order, although the
manipulation of terms is obviously more complicated.

We will only consider scalar perturbations in the following. Whereas
at linear order in perturbation theory, scalar, vector, and tensor
perturbations decouple, this is no longer the case at second order
(see e.g.~Ref.~\cite{MW2008}). However, since the amplitude of vector
and tensor perturbations is in general much smaller than that of the
scalars, we will still capture the dominant features of the theory,
incurring only a small error. We will revisit this issue in a future
publication.

\newpage
\subsection{Second-order equations}

The energy constraint at second order in a non-specific gauge form
is~\cite{christopherson:thesis},

\begin{align}
&3\H(\iipsi'+\H\iiphi)+\nabla^2\Big(\H(\BB-\EE')-\iipsi\Big)
+\nabla^2\B\Big(\nabla^2(\E'-\frac{1}{2}\B)-2\ipsi'\Big)\nn\\
&+\B_{,i}\Big(\H(3\H\B_,{}^i-2\nabla^2\E_,{}^i-2(\ipsi+\iphi)_,{}^i)
-2\ipsi'_,{}^i\Big)+2\E_,{}^{ij}(\ipsi-2\H\B)_{,ij}\nn\\
&+4\H(\ipsi-\iphi)\Big(3\ipsi'-\nabla^2(\E'-\B)\Big)
+\E'_,{}^{ij}\Big(4\H\E+\frac{1}{2}\E'-\B\Big)_{,ij}\nn\\
&+\ipsi'\Big(2\nabla^2(\E'-2\H\E)-3\ipsi')\Big)
+\ipsi_,{}^i(2\nabla^2\E-3\ipsi)_{,i}
+2\nabla^2\ipsi(\nabla^2\E-4\ipsi)\nn\\
&-12\H^2\iphi^2
+\frac{1}{2}\Big(\B_{,ij}\B_,{}^{ij}+\nabla^2\E_{,j}\nabla^2\E_,{}^j
-\E_{,ijk}\E_,{}^{ijk}-\nabla^2\E'\nabla^2\E'\Big)
\nn\\
&=-4\pi Ga^2\Big(2(\rhob+\Pb)V_{1,}{}^kv_{1,k}+\drhorho\Big)\,,
\label{en2:const}
\end{align}

\ni while the momentum constraint is
\begin{align}
&\iipsi_{,i}'+\H\iiphi_{,i}-\E_{,ij}'(\ipsi+\iphi+\nabla^2\E)_,{}^j
+\B_{,ij}(2\H\B+\iphi)_,{}^j\nn\\
&-\Big[\ipsi_{,i}(\nabla^2\E-4\ipsi)\Big]'
-\iphi_{,i}\Big(8\H\iphi+2\ipsi'+\nabla^2(\E'-\B)\Big)\nn\\
&-\B_{,j}\ipsi_{,i}{}^j+2\ipsi'_,{}^j\E_{,ij}+\E'_{,jk}{}\E_{,i}{}^{jk}
-\ipsi_{,i}'(\nabla^2\E+4\iphi)-\nabla^2\ipsi\B_{,i}\nn\\
&=-4\pi Ga^2\Big[(\rhob+\Pb)\Big(V_{2,i}-2\iphi(V_1+\B)_{,i}
-4(\ipsi v_{,i}-\E_{,ik}v_{1,}{}^k)\Big)\nn\\
&\qquad\qquad\qquad+2(\drho+\dP)V_{1,i}\Big]\,.
\label{mom2:const}
\end{align}

\ni These equations simplify if we specify a particular gauge. We
  choose the longitudinal gauge, as in the first-order analysis above
  (this gauge is extended to the Poisson gauge when vectors and
  tensors are included and also subjected to the shear-free
  gauge condition). In the longitudinal gauge Eq.~\eqref{en2:const}
  takes the form

\begin{align}
&3\H(\iipsi_{\lg}'+\H\iiphi_{\lg})- \nabla^2 \iipsi_{\lg}
-3 \phi_{1\ell}'^2 - 3 \phi_{1\ell,k}\phi_{1\ell,}{}^{k} - 8
\phi_{1\ell}\n^2\phi_{1\ell} - 12 \H^2 \phi_{1\ell}^2\nn 
\\
&=-4\pi Ga^2\rhob\Big(\delta\rho_{2\lg} + 2(1 + w)v_{1\lg,}{}^kv_{1\lg,k}\Big)\,,
\label{en2:long}
\end{align}

\ni while Eq.~\eqref{mom2:const} is reduced to

\begin{align}
  &\left(\psi_{2\lg}' + \H \phi_{2\lg}\right)_{,i} 
  + 2 \left(\phi_{1\ell,i}\phi_{1\ell}\right)' - 
  8 \H \phi_{1\ell,i}\phi_{1\ell} - 2 \phi_{1\ell,i}' \phi_{1\ell} \nn\\
  & \,= -4\pi Ga^2\rhob\Big\{
  (1 + w)\left[v_{2\lg,i} - 6  v_{1\lg,i}\phi_{1\ell} \right] 
  + 2(1 + c_{\rm s}^2)v_{1\lg,i}\delta_{1\lg} + 2 \frac{1}{\rhob}\delta
  P_{\rm nad 1}v_{1\ell, i}\Big\}\,.
  \label{mom2:long}
\end{align}

\ni Here $w = P_0 / \rhob$ is the equation of state of the fluid,  and
the non-adiabatic pressure perturbation, $\delta P_{\rm nad 1}$, is defined as
\be 
\delta P_{\rm nad 1}=\delta P_1 - c_{\rm s}^2 \delta\rho_1\,.
\ee 

Following the steps of the procedure at first order, we take the 
spatial divergence of Eq.~\eqref{mom2:long} and integrate with
the inverse Laplacian operator $\n^{-2}$. We obtain
\begin{align} 
\psi_{2\lg}'& + \H \phi_{2\lg} + \left(\phi_{1\ell}^2\right)'
 - 4 \H \phi_{1\ell}^2  - 2 \n^{-2}
 \left(\phi_{1\ell,j}'\phi_{1\ell}\right)_{,}{}^{j} \nn\\ 
 =& -4\pi G a^2\rhob \Big\{(1 + w) 
 \left[v_{2\lg} - 6\n^{-2} \left(
   v_{1\lg,j}\phi_{1\ell}\right)_{,}{}^{j} \right] + 2(1 + 
 c_{\rm s}^2)\n^2 \left(v_{1\lg,j}\delta_{1\lg}\right)_{,}{}^{j}  
+ 2 \frac{1}{\rhob}\left(\delta P_{\rm nad 1}v_{1\ell, j}\right)_{,}{}^{j}\Big\}.
\label{mom2:longint}
\end{align}

\ni We can now substitute this into Eq.~\eqref{en2:long} to
arrive at
\begin{align}
  \n^2 \psi_{2\lg}& +    \frac{3}{2} \n^2(\phi_{1\ell}^2) + 
  3 \left(\phi_{1\ell}'\right)^2 
  + 5 \phi_{1\ell}\n^2 \phi_{1\ell} + 3 \H \left(\phi_{1\ell}^2\right)' 
  - 6 \H \n^2 \left[ \phi_{1\ell,j}'\phi_{1\ell}\right]_{,}^{j}
  \label{poisson2:long} \\
  =& 4 \pi G a^2 \rhob \Big\{ \delta_{2\lg} - 3 \H (1 + w)v_{2\lg} + 2
  (1 + w) v_{1\lg,j}v_{1\lg,}{}^{j}  + 6 \H \n^{-2}\left[v_{,j} \left(3(1 + w)
    \phi_{1\ell} - (1 + c_{\rm s}^2) \delta_{1\lg} -
    \frac{1}{\rhob}\delta P_{\rm nad 1}\right)\right]_,^{~j}\Big\} 
  \nn
\end{align}

\subsection{The Poisson equation at second order}

To write the second-order equivalent of the Poisson equation in
\eqref{poisson1:tom}, we must transform the density contrast
to the total matter gauge. The transformation rule at second order is
given in \citep[][Eq.~(6.20)]{MW2008}   
\be
\wt{\delta \rho_{2\tom}} = \delta \rho_{2\lg} + \rhob'
\alpha_{2\tom}  + \alpha_{1\tom}\left(\rhob'' \alpha_{1\tom} + \rhob'
\alpha_{1\tom}' + 2 \delta\rho_{1\lg}'\right).
\ee
The second-order $\alpha_{2\tom}$ evaluated in the longitudinal gauge
is found with the aid of expression~(2.100) in
Ref.~\cite{christopherson:thesis} and using
Eqs.~\eqref{alpha1:tom}~and~\eqref{beta1:tom}. We obtain
\be
\alpha_{2\tom} = v_{2\lg} - \H v_{1\lg}^2 + \frac{1}{2}(v_{1\lg}^2)' 
- 4 \n^2 \left[v_{1\lg,j}\left(\phi_{1\ell} -
  v_{1\lg}'\right)\right]_{,}^{~j}.
\label{alpha2:tom}
\ee 

\noindent With the aid of the background equations and the expressions
for $\alpha_{1\tom}$ in Eq.~\eqref{alpha1:tom} and $\delta_{1\tom}$ in
Eq.~\eqref{delta1:tom} we obtain the gauge transformation,
\begin{align}
\wt{\delta_{2\tom}} =\,&\delta_{2\lg} - 3 \H (1 + w) v_{2\lg} - 3 \H (1 +
w)v_{1\lg}\delta_{1 \tom} + 2 v_{1\lg}\delta_{1\tom}' \nn\\
& + 12\H(1 + w)\n^{-2}\left[v_{1\lg,j} \left(\phi_{1\ell} +
  v_{1\lg}'\right)\right]_{,}^{~j} 
+ \left[3 \H w' -\frac{3}{2} \H^2 (1 + w)\left(5  + 9
  w\right)\right]v_{1\ell}^2\,. 
\label{delta2:tom}
\end{align}

We substitute the $\delta_{\lg}$ factors at both orders into 
Eq.~\eqref{poisson2:long} for the total matter gauge equivalents. The
final expression in terms of $\delta_{\tom},\, \phi_{1\ell}$ and
$v_{1\lg}$ is then 
\begin{align}
  \n^2 \psi_{2\lg}& + \frac{3}{2}\n^2(\phi_{1\ell}^2)  
  + 3 \left(\phi_{1\ell}'\right)^2 
  + 5 \phi_{1\ell}\n^2 \phi_{1\ell} + 3 \H \left(\phi_{1\ell}^2\right)' 
  - 6 \H \n^{-2} \left[\phi_{1\ell} \phi_{1\ell,j}'\right]_{,}{}^{j}\, \nn\\
  &=4\pi G a^2 \rhob \Bigg\{ \delta_{2\tom} + 6\H (1 + w) v_{1\lg}
  \delta_{1\tom} - 2 v_{1\ell}\delta_{1\tom}' + 
  2 (1 + w) v_{1\lg,j}v_{1\lg,}{}^{j} + \frac{3}{2} \H^2(1 + w)\left(3
  w - 1\right) v_{1\lg}^2 \nn\\ 
  & + 6 \H (1 + w)\n^{-2} \left[v_{1\lg,j}\left(\phi_{1\ell} - 2 v_{1\lg}' -
    \left(\frac{1+ c_{\rm s}^2}{1 + w}\right) 
    \delta_{1 \tom} - \frac{1}{1 + w} 
    \frac{\delta P_{\rm nad 1}}{\rhob}\right)\right]_{,}^{~j} 
  \Bigg\}.    
  \label{poisson2:tom}
\end{align}
  
\ni This rather long equation fulfils our first goal, to provide a
Poisson equation at second order using the same variables employed in
the structure formation studies at the Newtonian limit. It is already
clear that adopting the expression $\nabla^2 \psi_2 = 4 \pi G a^2
\rhob \delta_2$ leaves out most of the terms of the actual second
order Poisson constraint. 

To conclude this section, let us rewrite Eq.~\eqref{poisson2:tom} in
terms of the potential $\phi_{2\ell}$ instead of $\psi_{2\ell}$. This
will come handy in the next section since primordial non-Gaussianity
is conventionally formulated in terms of this variable. We use the
traceless $ij$ component of the field equations, derived from the
Eq.~(A.1) in \cite{christopherson:thesis}, written in the longitudinal
gauge as 
\begin{align}
\n^4 (\psi_{2\ell} - \phi_{2\ell}) =& - 4 \n^4 (\phi_{1\ell}^2) + 2
\phi_{1\ell,j}{}^i \phi_{1\ell,i}{}^j + 6
\n^2 \phi_{1\ell}\n^2 \phi_{1\ell} +  8  \phi_{1\ell,j}\n^2
\phi_{1\ell,}{}^{j} \notag \\
+ 4 \pi G a^2 \rhob &\left\{2 (1 + w) 
\left[\n^2 (v_{1\ell,j}v_{1\ell,}{}^{j}) + 3 \n^2(v_{1\ell}\n^2v_{1\ell}) + 3
  \n^2(v_{1\ell})\n^2(v_{1\ell}) - 3 v_{1\ell}\n^4 v_{1\ell}\right]\right\}\,.
\label{psi2:phi2}
\end{align}
 
\noindent Upon substitution of this in the constraint equation,
Eq.~\eqref{poisson2:tom}, and with some algebra we arrive at  
\begin{align}
&\n^4 \phi_{2 \ell} - 2 \n^4 (\phi_{1\ell}^2) + 7
  \n^2(\phi_{1\ell}\n^2\phi_{1\ell})  + 3 (\n^2\phi_{1\ell})^2 
- 3 \phi_{1\ell} \n^4 \phi_{1\ell} + 3 \n^2(\phi_{1\ell}'{}^2)  + 
+ 3 \H \n^2\phi_{1\ell}'
- 6 \H \left(\phi_{1\ell,j}'\phi_{1\ell}\right)_,{}^j\,  \nn\\  
&= 4 \pi G a^2 \rhob  \Bigg\{ \n^2 \delta_{2 \tom} + 6 \H (1 + w) \n^2
(v_{1\ell}\delta_{1\tom}) - 2 \n^2 (v_{1\ell}\delta_{1\tom}') 
+ \frac32 \H^2 (1+w)(3w -1) \n^2 v_{1\ell}^2  \nn\\ 
& \,+ 6 (1 + w)\left[v_{1\ell} \n^4 v_{1\ell} -\n^2(v_{1\ell}\n^2v_{1\ell}) 
  - (\n^2v_{1\ell})^2 
+ \H \left(v_{1\ell,j} \left(\phi_{1\ell} - 2 v_{1\ell}' -
  \frac{1 + c_{\rm s}^2}{1 + w} \delta_{1\tom} - \frac{1}{1+w}
  \frac{\delta P_{\rm nad 1}}{\rhob}\right)\right)_,^{~j} \right]
\Bigg\}.
\label{poisson2:complete}
\end{align}

This constraint is valid for any perfect fluid. In the following
section we show how to insert this constraint in the initial
conditions of numerical simulations of structure formation.

\section{Non-Gaussian initial conditions for numerical simulations}
\label{sec:initial:conditions}

The Newtonian Poisson equation is used at all orders as a constraint
to the initial conditions in numerical simulations. However, 
the above constraint is the one that provides consistency with General
Relativity. Imposed at an initial time, this constraint is met at all
times if the perturbations are evolved in the context of GR. It is
therefore useful to write the expression we have derived in terms of
variables employed in numerical simulations, namely  
$\delta_{1\tom}$ and $ v_{1\lg}$, evaluated at some initial time.
Here we derive such an expression with the aid of the first order
equations, Eqs.~\eqref{mom:constlg}, \eqref{poisson1:tom}, and the continuity
equation from Ref.~\cite{Christopherson:2012kw} in terms of the chosen gauge. These
help us to replace the time derivatives in the constraint
equations. After some more algebra we obtain
\begin{align}
\n^4 \phi_{2\lg} & - 2 \n^4 (\phi_{1\ell}^2) + 7 \n^2
(\phi_{1\ell}\n^2 \phi_{1\ell}) + 3(\n \phi_{1\ell})^2 -
3\phi_{1\ell}\n^4 \phi_{1 \ell}    \nn\\
&= 4 \pi G a^2\rhob \Bigg\{\n^2 \delta_{2\tom} 
+ 6 (1+w)\left[v_{1\ell}\n^4 v_{1\ell}
  -(\n^2v_{1\ell})^2 - \frac23 \n^2(v_{1\ell} \n^2 v_{1\ell}) +
  \frac34 (1+w)\H^2 \n^2 (v^2_{1\ell})\right] \nn\\
& +  6 \H \n^2 (v_{1\ell}\delta_{1\ell}) + 6(1+w)\H
\left[v_{1\ell,j}\left(2\phi_{1\ell}
  -\frac{c_s^2-1}{1+w}\delta_{1\tom} 
  + \frac{1}{1+w}\frac{\delta P_{\rm nad 1}}{\rhob}\right) 
  \right]_,^{~j} \Bigg \}.
\label{poisson2:initial}
\end{align}

We emphasise that the {Newtonian} counterpart of this constraint is a
linear equation which includes only the first term at each side of the
equality. All the other terms bring relativistic contributions to the
Poisson equation. This expression can be used in the numerical
simulations that set initial conditions for perturbations of in any
perfect fluid and allowing for entropy perturbations.

To reduce Eq.~\eqref{poisson2:initial} further, we can either
eliminate the density contrast $\delta_{1\tom}$ or the potential 
$\phi_{1\ell}$ via the first-order Poisson equation
\eqref{poisson1:tom}. This proves useful when we want to make contact
with formulations like the so-called renormalised perturbation theory
(RPT)\cite{Crocce:2005xy}, where the initial conditions are set,
order by order in Fourier space, via recursive relations in powers of
$\delta_{1\New}$ (see, e.g., Ref.~\cite{Bernardeau:2001qr}). 
To reduce things further, let us focus on the case of an
Einstein-de Sitter universe, a flat space-time filled by dust, i.e., where  
$w =0$ as well as $\delta P_1 = 0$ and $\Lambda = 0$. In this case,
Eq.~\eqref{poisson2:initial} is reduced to  
\begin{align}
\n^4 \phi_{2\lg} & - 2 \n^4 (\phi_{1\ell}^2) + 7 \n^2
\phi_{1\ell}\n^2 \phi_{1\ell} - 3\phi_{1\ell}\n^4 \phi_{1 \ell}  + 3
(\n^2\phi_{1\ell})^2   \nn\\
&= 4 \pi G \rhob\Big\{\n^2 \delta_{2\tom} 
 + \frac92 \H^2 \n^2 v_{\ell}^2 
+ 2\left[3 v_{1\ell}\n^4 v_{1\ell} - 3 (\n^2v_{1\ell})^2 - 
2 \n^2 (v_{1\ell}\n^2v_{1\ell})\right]  \nn \\
& \qquad \quad + 6\H \n^2
(v_{1\ell}\delta_{1\ell}) + 6 \H \left[v_{1\ell,j}\left(2 \phi_{1\ell} -
  \delta_{1\tom}\right)\right]_,^{~j}   
\Big \}.
\label{poisson2:dust}
\end{align}

\noindent With the aim of incorporating our result as an
initial constraint in the formulation of non-linear initial conditions
for numerical simulations, we transform Eq.~\eqref{poisson2:dust} to
the Fourier space. Additionally, as is customary in structure
formation studies, we work exclusively with the growing
mode of perturbations, where $\phi_{1\ell} = \mathrm{const}$. It is
then possible to write all of first order variables in terms of
$\delta_{1\tom}$ (as in the standard 
perturbation theory, c.f. Ref.~\cite{Bernardeau:2001qr}) with the aid of the
first order Poisson equation \eqref{poisson1:tom} and the momentum constraint
at first order. Explicitly, in Fourier space,
\be
{\phi_{1\ell}}(k) = - \frac32 \frac{\H^2}{k^2} {\delta_{1\tom}}(k),
\qquad {v_{1\ell}}(k) = \frac{\H}{k^2} {\delta_{1\tom}}(k).
\ee

\noindent The second relation above follows directly from
Eq.~\eqref{mom:constlg} and the first equivalence above, keeping in
mind that we are working with the growing mode exclusively.   
The reduced Poisson equation at second order is
\begin{align}
  k^4 {\phi_{2\ell}}(k)& + \frac32 k^2\H^2
  {\delta_{2\tom}}(k) =  \frac32 \int\,d^3 p d^3 q
  \delta_{D}^3 (p + q - k)\frac{\H^4}{p^2q^2}
  \nn\\ 
  &  \times \bigg\{3 {|p + q|^4} + \frac{15}{4}(p^4 + q^4)  
  - \frac{35}{4}|p+q|^2 (p^2 + q^2) - \frac{15}{2} p^2q^2  
  + \frac92 \H^2|p + q|^2 \bigg\} {\delta_{1\tom}}(p) 
  {\delta_{1\tom}}(q)\,, 
  \label{poisson2:fourier}
\end{align}

\noindent where $\delta_{D}(k)$ is the Dirac delta function and where
$k$-modes in the integral are represented by $k = |p + q|$. 
This equation represents a concrete constraint for initial conditions of
numerical simulations, consistent with GR, and written in terms of
relativistic equivalents to the gravitational potential and the matter
density perturbation. Note that, while the linear equation of
Ref.~\cite{Hwang:2012bi} is valid for these second order variables  
at small scales, the relativistic corrections obtained here become
increasingly important as the perturbation modes approach the horizon
scale.

Since this constraint already carries couplings between different
perturbation modes, there will be some \emph{intrinsic}
non-Gaussianity induced by this second-order correspondence. This is a
known effect of GR which has recently been explored in the CMB through the use
of second order Boltzmann codes
\cite{Huang:2012ub,Su:2012gt,Pettinari:2013he}, and in the matter
density field \cite{Bartolo:2010rw}. Here we disentangle 
the effect of the initial constraint from the influence of the non-linear
evolution of perturbations.
To observe the type of non-Gaussianity induced by the GR constraint, we
introduce three templates that constitute a basis for the non-Gaussian
$\phi_{2\ell}$. These templates are also a basis to represent the
initial conditions of the density contrast with primordial
non-Gaussianity, and consistent with GR, for a given model of
structure formation.   

Following the convention of \cite{Komatsu:2001rj} for the
non-Gaussianity in the lapse function, the local template is, 
\be
\phi_{\ell}^{loc} = \phi_{\ell G} + f_{\rm NL}^{loc}\left[ \phi_{\ell G}^2 -
  \langle\phi_{\ell G}^2\rangle\right].
\label{local:prescription}
\ee

\noindent This is preserved in super-horizon scales since $\phi_{\ell}$ is
constant when the universe is filled with dust. Therefore, we can
directly substitute the primordial $\phi_{2\ell}$ in the Poisson constraint
\eqref{poisson2:fourier}.   
  
In Fourier space, the local configuration in
Eq.~\eqref{local:prescription} yields
\be
\frac12 {\phi_{2\ell}^{loc}}(k) = 
\frac{9}{4}f_{\rm NL}^{loc}\int\,d^3 p d^3 q
\delta_{D}^3 (p + q - k)\left(\frac{\H^4}{p^2q^2}\right)
{\delta_{1\tom}}(p) {\delta_{1\tom}}(q)\,,
\label{phi2:local}
\ee

\noindent and we can generate a kernel for $\delta_{2\tom}^{loc}$,
\begin{align}
  \frac{k^2}{\H^2} &
  \delta_{2\tom}^{loc}(k) =  \int\,d^3 p d^3 q
  \frac{1}{p^2q^2}\delta_{D}^3 (p + q - k)
  \nn\\  
  &  \times \bigg\{3\left(1 - f_{\rm NL}^{loc}\right) {|p + q|^4}
  + \frac{15}{4}(p^4 + q^4)
  - \frac{35}{4}|p+q|^2 (p^2 + q^2) - \frac{15}{2} p^2q^2 + 
  \frac92 \H^2|p + q|^2 \bigg\} \delta_{1\tom}(p)
  \delta_{1\tom}(q)\,.
  \label{delta2:local}         
\end{align}

Note that the primordial non-Gaussianity of the local configuration
has the same momentum dependence as one of the terms if the
relativistic constraint in Eq.~\eqref{poisson2:fourier}. This is shown explicitly
in the last equation and we can interpret this as an intrinsic
relativistic contribution to the non-Gaussianity observable in the
Large scale structure. We denote this GR contribution as 
$f_{\rm NL}^{loc (GR)}$ with a numerical value $ f_{\rm NL}^{loc (GR)} = -1$. 
Repeating the procedure for the equilateral and orthogonal
configurations, we can provide initial conditions for
$\delta_{2\tom}$ in a complete basis for primordial non-Gaussian
perturbations. We borrow the templates implemented in 
Ref.~\cite{Scoccimarro:2011pz}. For the equilateral
configuration, this template is   
\begin{align}
  \frac12 k^4 \phi_{2\ell}^{eq}(k) &= \frac94 f_{\rm NL}^{eq}\int\,d^3 p d^3 q
  \delta_{D}^3 (p + q - k)\left(\frac{\H^4}{p^2q^2}\right)\delta_{1\tom}(p)
  \delta_{1\tom}(q)\nn \\ 
  &\times\left\{  - 3 |p + q|^4 + (p^2 - q^2) |p + q|^{2} + 2
  (p + q)|p + q|^{3}  \right\}  \,,  
\label{phi2:equilateral}
\end{align}

\noindent while in the orthogonal case
\begin{align}
  \frac12 k^4 \phi_{2\ell}^{ort}(k) &= \frac94 f_{\rm NL}^{ort}\int\,d^3 p d^3 q
  \delta_{D}^3 (p + q - k)\left(\frac{\H^4}{p^2q^2}\right)\delta_{1\tom}(p)
  \delta_{1\tom}(q)\nn \\ 
  &\times\left\{  - 9 |p + q|^4 + 4 (p^2 - q^2) |p + q|^{2}  + {5}
  (p + q)|p + q|^{3} \right\}  \,.  
  \label{phi2:orthogonal}
\end{align}

Finally the complementary equilateral and orthogonal kernels for
$\delta_{2\tom}$ are   
\begin{align}
  \frac{k^2}{\H^2}&
  {\delta_{2\tom}^{eq}}(k) =  \int\,d^3 p d^3 q
  \frac{1}{p^2q^2}\delta_{D}^3 (p + q - k)
  \nn\\  
  &  \times \Bigg\{3\left(1 + 3 f_{\rm NL}^{eq}\right) {|p + q|^4}  
- \frac14 \left(12 f_{\rm NL}^{eq}  + 35\right) |p+q|^2 (p^2 + q^2) 
+ 6 f_{\rm NL}^{eq}\left(|p+q|^2 pq
- |p + q|^3 (p+q)\right) \nn \\ 
  &\quad  + \frac{15}{4}(p^4 + q^4)
   - \frac{15}{2} p^2q^2 + 
  \frac92 \H^2|p + q|^2 
  \Bigg\} {\delta_{1\tom}}(p) {\delta_{1\tom}}(q),
  \label{delta2:equilateral}         \\
  \frac{k^2}{\H^2}&
  {\delta_{2\tom}^{ort}}(k) =  \int\,d^3 p d^3 q
  \frac{1}{p^2q^2}\delta_{D}^3 (p + q - k)
  \nn\\  
  & \times \Bigg\{3\left(1 + 9 f_{\rm NL}^{ort}\right) {|p + q|^4}  
- \frac14 \left( 48 f_{\rm NL}^{ort} + 35\right) |p+q|^2 (p^2 + q^2)
+ 3 f_{\rm NL}^{ort} \left(8 |p+q|^2 pq  
- 5 |p + q|^3 (p+q) \right) \nn \\ 
  &\quad  + \frac{15}{4}(p^4 + q^4)
   - \frac{15}{2} p^2q^2 + \frac92 \H^2|p + q|^2 
  \Bigg\} 
    {\delta_{1\tom}}(p)   {\delta_{1\tom}}(q).
  \label{delta2:orthogonal}           
\end{align}

These three kernels represent a complete basis for the primordial
bispectrum. The kernels above show that the Poisson constraint yields 
different contributions for the local, equilateral and orthogonal
configurations. The relativistic initial conditions can mimic non-Gaussian
contributions as discussed after Eq.~\eqref{delta2:local}. This
intrinsic relativistic non-Gaussian imprint in the matter fluctuation
has a value in the local template of $f_{NL}^{loc(GR)} = -1$ (this is 
particularly relevant in studies of LSS since it is the dominant
configuration contributing to the halo bias \cite{Matarrese:2000iz,
  Dias:2013rla}).  

In the equilateral configuration, we can read the intrinsic GR
contributions to $f_{\rm NL}$
from the parentheses in Eq.~\eqref{delta2:equilateral}. The dominant
contribution is $f_{\rm NL}^{eq(GR)} = 35/12$, while for the orthogonal
configuration, we read from Eq.~\eqref{delta2:orthogonal} $f_{\rm
  NL}^{ort (GR)} = 35/48$ as a dominant contribution. Note that, although
there is an extra contribution of GR to $f_{\rm NL}^{(GR)}$ in each
one of these configurations, we quote the largest numerical value for
each case.  
 
The result obtained for the intrinsic non-Gaussianity in the local
configuration {is compared with that of Ref.~\cite{Bartolo:2010rw}, 
for the equivalent case of the Poisson gauge, in the appendix \ref{app:a}.} 
A detailed analysis of the modification of $f_{\rm NL} $ separating 
initial constraints from the non-linear evolution of $\delta_{2}$ in
the synchronous-comoving gauge is the subject of a recent paper 
\cite{bruni:2013}. For our purposes, it suffices to emphasise that the
results of this section are written in terms of the GR variables that
find a direct correspondence with the Newtonian ones, since we intend
to present initial conditions for the numerical studies of galaxy
formation.  

\section{Discussion}
\label{sec:discussion}

In this paper we have derived the relationship between the energy density 
fluctuation and the curvature perturbation at second order in the context of 
cosmological perturbation theory. This Poisson equation at second
order, presented in Eq.~(\ref{poisson2:initial}) in full generality
for a single fluid including entropy perturbations, is expressed in
terms of variables equivalent to an Eulerian set in Newtonian hydrodynamics.
We found that the Poisson equation takes a particularly simple form at
second order if the matter density fluctuation is expressed in the total matter
gauge, and not the comoving orthogonal gauge which has been used
before at first order. For the Poisson equation, the difference of the
two gauges only becomes apparent at second order in perturbation theory.

As an example, we calculate the second order Poisson equation in the
case of an Einstein-de Sitter universe, and present the result in
Eq.~\eqref{poisson2:fourier} in Fourier space. We show
how to incorporate primordial non-Gaussianity into the matter
perturbation at second order in an equation consistent with GR. In
this way, we can also quantify the non-Gaussianity intrinsic to GR
contributions. Our results generalise the non-Gaussian kernels
presented in terms of Newtonian physics in
Ref.~\cite{Scoccimarro:2011pz} to include relativistic terms.  We show
that the non-linearity of GR induces a non-Gaussian signature in
addition to the primordial value. In particular we find, in the local
configuration, a value $f_{NL}^{loc(GR)} = -1$, consistent with that
obtained in Ref.~\cite{Bartolo:2010rw} in the Poisson gauge.

Achieving consistency with the result of \cite{Bartolo:2010rw}
in this limit shows the strength of our results since we can recover
the primordial and the GR contribution to non-Gaussianity in
$\delta_2(\tau,{\bf x})$ in 
a $\Lambda$-CDM universe without solving the field equations. 
Our result, the Poisson equation at second order, and the example
presented in this paper provide fairly simple equations that can be
directly incorporated into generators of initial conditions for  
numerical simulations which take care of the evolution of
fluctuations. The initial conditions generated in this way 
account for general relativistic effects in N-body codes and other
numerical simulations of structure formation.

\section*{Acknowledgements}
JCH is grateful to Marc Manera for useful discussions. 
The authors are grateful for the support of the DGAPA-UNAM through the
grant PAPIIT IN116210-3. AJC acknowledges support from the European
Commission's Framework Programme 7, through the Marie Curie
International Research Staff Exchange Scheme LACEGAL
(PIRES-GA​-2010-2692​64) and is grateful to the IA-UNAM, ICN-UNAM, and
QMUL for hospitality.     
JCH is funded by CONACYT (CVU No. 46280), AJC by the Sir Norman
Lockyer Fellowship of the Royal Astronomical Society, and KAM is 
supported, in part, by STFC grant ST/J001546/1.
\appendix

\section{Comparison with previous work}
\label{app:a}
\numberwithin{equation}{section}

In this appendix we show that our result is consistent with that
reported in \cite{Bartolo:2010rw} at the level of initial conditions, 
and that the difference at face is only due to the definitions used in that paper. 

Let us first note that the transformations performed to change the
matter variable $\delta_{2 \ell}$ to $\delta_{2\tom}$ do not modify the
curvature sector of the Poisson equation as can be seen from comparing
Eqs.~\eqref{poisson2:long} and~\eqref{poisson2:tom}. The only
modification to the curvature dependence is due to the change of
variables from $\psi_{2\ell}$ to  $\phi_{2\ell}$. The relevant terms
that determine the local configuration of non-Gaussianity and its GR
correction are, from Eq.~\eqref{poisson2:complete},
\be
\n^4 \phi_{2 \ell} - 2 \n^4 (\phi_{1\ell}^2) + \ldots
\ee

Note that the analysis of Sec.~\ref{sec:initial:conditions} does not
modify these terms and ultimately, using the definition
\eqref{local:prescription}, this second term is responsible for
the GR induced non-Gaussianity, which yields $\fNL^{loc (GR)} = -1$. 

The above argument shows that the same value for induced
non-Gaussianity is recovered when we work with all variables in the
longitudinal gauge. The corresponding result in \cite{Bartolo:2010rw}
is derived from the first term of Eq.~(31), that is, 
\be
\fNL^P  \supset \left[\frac53 (a_{\rm NL} -1) + 1 - \frac{g(\tau)}{g_{in}} -
  \frac12 \frac{B_{1}(\tau)}{g(\tau)g_{in}}\right],
\label{fnl:poisson}
\ee

\noindent where the functions of time have an explicit argument and
play no role in the initial conditions. The parameter $a_{\rm NL}$ is
defined in terms of the curvature perturbation in uniform density
hypersurfaces by the equivalence $\zeta_2 = 2 \zeta_1^2$. Subsequently,
the definition of the parameter $f_{\rm NL}^{LR}$ in
Ref.~\cite{Lyth:2005du} indicates that  
\be
\zeta_2 = \left(\frac65 \fNL^{LR} + 2\right)\zeta_1^2  \qquad
\Rightarrow \quad a_{\rm NL} -1 = \frac35 \fNL^{LR}.
\label{anl:fnl}
\ee

This definition is not stated explicitly in \cite{Bartolo:2010rw}
but it is implied by the limits discussed at the end of Sec.~2 of that
paper. The primordial non-Gaussianity in this case is thus given by
$\fNL^{LR}$. We can then subtract the primordial non-Gaussianity from
Eq.~\eqref{fnl:poisson} and ignore the time-dependent part to find that 
\be
\fNL^{P (GR)} = 1.
\ee  

Let us finally note that the parameter $\fNL^{P}$ is constructed from the
definitions in Eqs.~(4) and~(5) of \cite{Bartolo:2010rw}. For our
variables this implies that 
\be
\phi_{2\ell} = - 2\fNL^P( \phi_{\ell G} - \phi_{\ell G}).
\ee 

In view of Eq.~\eqref{local:prescription} we find the equivalence $\fNL^P = -
\fNL^{loc}$ and thus recover the result obtained in the body of the
paper.

\end{document}